# Digital terrain modeling with the Chebyshev polynomials

I.V. Florinsky[*], A.N. Pankratov

Institute of Mathematical Problems of Biology, Russian Academy of Sciences
Pushchino, Moscow Region, 142290, Russia

**Abstract**

Mathematical problems of digital terrain analysis include interpolation of digital elevation models (DEMs), DEM generalization and denoising, and computation of morphometric variables by calculation of partial derivatives of elevation. Traditionally, these procedures are based on numerical treatments of two-variable discrete functions of elevation. We developed a spectral analytical method and algorithm based on high-order orthogonal expansions using the Chebyshev polynomials of the first kind with the subsequent Fejér summation. The method and algorithm are intended for DEM analytical treatment, such as, DEM global approximation, denoising, and generalization as well as computation of morphometric variables by analytical calculation of partial derivatives. To test the method and algorithm, we used a DEM of the Northern Andes including 230,880 points (the elevation matrix $480 \times 481$). DEMs were reconstructed with 480, 240, 120, 60, and 30 expansion coefficients. The first and second partial derivatives of elevation were analytically calculated from the reconstructed DEMs. Models of horizontal curvature ($k_h$) were then computed with the derivatives. A set of elevation and $k_h$ maps related to different number of expansion coefficients well illustrates data generalization effects, denoising, and removal of artifacts contained in the original DEM. The test results demonstrated a good performance of the developed method and algorithm. They can be utilized as a universal tool for analytical treatment in digital terrain modeling.

## 1. Introduction

Topography is one of the main factors controlling processes taking place in the near-surface layer of the planet. In particular, topography is one of the soil forming factors since it influences: (a) climatic and meteorological characteristics, which controls hydrological and thermal regimes of soils; (b) prerequisites for gravity-driven overland and intrasoil lateral transport of water and other substances; and (c) spatial distribution of vegetation cover. At the same time, being a result of the interaction of endogenous and exogenous processes of different scales, topography can reflect the geological structure of a terrain. In this connection, digital terrain analysis and digital terrain models (DTMs) are widely used to solve various multiscale problems of geomorphology, hydrology, remote sensing, soil science, geology, geophysics, geobotany, glaciology, oceanology, climatology, planetology, and other disciplines (Wilson and Gallant, 2000; Li et al., 2005; Hengl and Reuter, 2009; Florinsky, 2012).

Mathematical issues of quantitative modeling and analysis of the topographic surface can be summarized in three main problems (Florinsky, 2012):

1. Interpolation of digital elevation models (DEMs), two-dimensional discrete functions of an elevation defining the topographic surface as a set of values measured at the grid nodes. This task is commonly carried out by local interpolation methods, such as piecewise splines, etc. (Watson, 1992).
2. DEM filtering to denoise, generalize, and decompose DEMs into components of

[*] Correspondence to: iflorinsky@yahoo.ca

different spatial scales. These tasks are usually attacked by two-dimensional (2D) discrete Fourier transform, 2D discrete wavelet transform, smoothing, row and column elimination from DTMs, and 2D singular spectrum analysis (Florinsky, 2012, pp. 103–132).

3. Derivation of morphometric variables from DEMs to describe and analyze geometric peculiarities of the topographic surface. If the topographic surface is defined by a continuous, single-valued function

$$z = f(x, y), \qquad (1)$$

where $z$ is elevation, $x$ and $y$ are the Cartesian coordinates, local topographic variables are functions of partial derivatives of elevation:

$$r = \frac{\partial^2 z}{\partial x^2}, \ t = \frac{\partial^2 z}{\partial y^2}, \ s = \frac{\partial^2 z}{\partial x \partial y}, \ p = \frac{\partial z}{\partial x}, \ q = \frac{\partial z}{\partial y}. \qquad (2)$$

For example, horizontal curvature ($k_h$), one of the most important morphometric attributes, is calculated by the following equation (Shary, 1995):

$$k_h = -\frac{q^2 r - 2pqs + p^2 t}{(p^2 + q^2)\sqrt{1 + p^2 + q^2}}, \qquad (4)$$

To compute $r$, $t$, $s$, $p$, and $q$ (2) from DEMs based on plane square grids or spheroidal equal angular grids, one can apply methods based on approximation of partial derivatives by finite differences using the 3 × 3 or 5 × 5 moving windows (Evans, 1979, p. 29; Zevenbergen and Thorne, 1987; Shary, 1995; Florinsky, 1998, 2009).

It is however obvious that the three problems of digital terrain analysis may be resolved within a framework of an analytical treatment of DEMs using global approximation with (orthogonal) polynomials. In the 1970–1980s, there were attempts to apply higher order polynomials for DEM global approximation (Van Rossel, 1972; Segu, 1985). However, these attempts have faced a number of limitations: First, such approaches have required considerable computer resources. Second, practical tasks have demanded to work with increasingly large DEMs including tens of thousands to millions of points; existed methods and computers could not handle such data amounts. Third, the topographic surface has appeared too complex for wide application of global polynomial approximations. At the same time, the progress in the theory and practice of polynomial approximation (Dedus et al., 1999, 2004; Tetuev and Dedus, 2007) suggest that it is now possible to solve such problems.

In this paper we describe a spectral analytical method and algorithm based on high-order orthogonal expansions using the Chebyshev polynomials of the first kind with the subsequent Fejér summation. The method and algorithm are intended for DEM analytical treatment, such as, DEM global approximation, denoising, and generalization as well as computation of morphometric variables by analytical calculation of partial derivatives.

## 2. Method

Let's consider a function of two variables (1) defined in a rectangular domain. To approximate analytically this function, we use the two-dimensional expansion by the orthogonal Chebyshev polynomials of the first kind

$$z = \sum_{i=0}^{l-1}\sum_{j=0}^{l-1} d_{ij} T_i(x) T_j(y), \qquad (4)$$

where $T_i(x)$, $T_j(y)$ are the Chebyshev polynomials orthogonal in the interval [−1, 1], $d_{ij}$ are expansion coefficients. It is assumed that the domain of the initial function is translated into the domain of the orthogonal polynomials by a linear transformation.

The method and algorithms for the function expansion by the Chebyshev polynomials are described in (Press et al., 1992, p. 190–194; Pankratov, 2004). To calculate expansion coefficients, we use an operator, or matrix method introduced and studied by Pankratov (2004). In this study, a two-dimensional approximation of the function (1) is performed as a superposition of one-dimensional approximations by each variable.

We use the following formula for the Chebyshev polynomials:

$$B = \left\{ T_0 = \frac{1}{\sqrt{2}}, T_i = \cos(i \arccos x) \right\}, \qquad (5)$$

where $i$ is a number of the basis function, $i = 0, \ldots, l-1$. The system of functions $T_i(x)$ satisfies the orthogonality condition in the scalar product defined as follows:

$$(T_i, T_j) = \int_{-1}^{1} T_i(x) T_j(x) \frac{1}{\sqrt{1-x^2}} dx = \begin{cases} \pi, & i = j \\ 0, & i \neq j \end{cases}. \qquad (6)$$

In the discrete form, the scalar product on a nonuniform grid of $k$ nodes

$$t_i = \cos\left(\frac{\pi(i - 1/2)}{k}\right), \qquad (7)$$

which are the zeros of the orthogonal polynomial $T_k(x)$, $i = 0, \ldots, k$, has the form:

$$(T_n, T_m) = \frac{2}{k} \sum_{i=1}^{k} T_n(t_i) T_m(t_i) = \begin{cases} \pi, & n = m \\ 0, & n \neq m \end{cases}. \qquad (8)$$

The expansion coefficients are calculated by the expressions:

$$c_i = \frac{(f, T_i)}{(T_i, T_i)}. \qquad (9)$$

Approximations based on orthogonal polynomials always leads to oscillatory artifacts due to the Gibbs phenomenon (Jerri, 1998). To solve this problem, we replace the original approximation with a smooth representation obtained as an arithmetic mean of all partial sums of an orthogonal series (the Fejér summation). According to the Fejér theorem (Courant and Hilbert, 1989, p. 102), the initial mean-square approximation becomes a uniform one in this representation. The Fejér summation (or averaging) is the powerful method to suppress or eliminate oscillatory artifacts of the Gibbs phenomenon in data sets approximated with polynomials (Jerri, 1998; Pankratov and Kulikova, 2006).

Transformation of expansion coefficients corresponding to the arithmetic mean of the partial sums of an orthogonal series has the form:

$$\tilde{c}_i = \frac{l-i}{l} c_i, \qquad (10)$$

where $\tilde{c}_i$ are new weighting coefficients of an orthogonal series, $i = 0, \ldots, l-1$.

After the approximation of the function $z$ (1) and the suppression of oscillatory artifacts, it is possible to calculate the partial derivatives $r$, $t$, $s$, $p$, and $q$ (2) in an analytical representation similar to the function $z$, that is, in the form of two-dimensional orthogonal series. For the case of the Chebyshev polynomials, calculation of the expansion coefficients of a derivative from the expansion coefficients of the initial function is described by Press et al. (1992, p. 195). In general form for an arbitrary basis, a coefficient conversion scheme was presented by Tetuev and Dedus (2007).

In our case, these formulas take the following form:

$$\begin{aligned} p_{l-1} &= 0, \\ p_{l-2} &= 2(l-2)c_{l-1}, \\ p_j &= p_{j+2} + 2j c_{j+1}, \\ p_0 &= \frac{p_0}{\sqrt{2}}, \end{aligned} \qquad (11)$$

where $p_j$ are expansion coefficients of the derivative of an orthogonal series, $j = l-3, \ldots, 0$.

It should also be noted that expansion coefficients of the derivative should be scaled in the case of a linear transformation of the function domain:

$$p_j = \frac{2}{T} p_j, \qquad (12)$$

where $T$ is a length of a function interval, $j = 0, \ldots, l-1$.

Finally, to derive local morphometric variables, all the calculated values of the derivatives are substituted into related equations, for instance, the $k_h$ equation (3).

### 3. Algorithm

Let the initial array is specified as a matrix $A$ with dimension $m \times n$ representing the function values at the nodes of a square grid; $x_1, \ldots, x_m$ are values of the grid nodes in the interval $[-1, 1]$ along abscissa, $y_1, \ldots, y_n$ are values of the grid nodes in the interval $[-1, 1]$ along ordinate; $t_1, \ldots, t_k$ are values of the Gaussian quadrature nodes.

$k$ values determine the maximum degree of expansion $l$ in (4). Since $m \approx n$, so the Gaussian grid is identical for abscissas and ordinates. For a more accurate calculation of the expansion coefficients, it is recommended to choose the $k$ value greater than $m$ and $n$. In this study $k = 8\max(m, n)$.

Let's introduce the following notations: $L_{xt}$ is a matrix of linear interpolation from the grid $x$ to the grid $t$; $T_t$ is a matrix of the Chebyshev polynomial values $T_i(t_j)$ in the grid $t$, where $i = 0, \ldots, l-1$; $j = 1, \ldots, k$; $F$ is a diagonal matrix with diagonal elements $(l-i)/l$, $i = 0, \ldots, l-1$ to derive the arithmetic mean of orthogonal series sums.

Calculation of the expansion coefficients is carried out in two stages. First, all columns of the matrix $A$ are transformed into the expansion coefficients that corresponds to the expansion in the variable $y$:

$$C = \frac{2}{k} FT_t L_{yt} A. \tag{13}$$

Then the matrix of the expansion coefficients is transposed and the expansion is repeated that corresponds to the expansion in the variable *x*:

$$D = \frac{2}{k} FT_t L_{xt} C^\mathsf{T}, \tag{14}$$

where *C* and *D* are matrices of the expansion coefficients, the intermediate and resulting ones.

Reconstruction of the approximated function is carried out by a simple summation of orthogonal series:

$$Z = T_x^\mathsf{T} D^\mathsf{T} T_y. \tag{15}$$

Let *E* be the differentiation operator in the space of expansion coefficients. Then the coefficient matrix corresponding to the partial derivatives *p*, *r*, *q*, *t*, and *s* (2) have the following form:

$$\begin{aligned} P &= ED, \\ R &= E^2 D, \\ Q &= DE^\mathsf{T}, \\ T &= D(E^\mathsf{T})^2, \\ S &= PE^\mathsf{T}. \end{aligned} \tag{16}$$

These formulas exist only if the matrices of expansion coefficients are square.

### 4. Data

To test the method and algorithm proposed, we selected a portion of the Northern Andes measuring 4º × 4º, located between 2ºS and 2ºN, and 78º30'W and 74º30'W. The area covers regions of Ecuador, Colombia, and Peru including parts of the Coastal plain, the Andean Range, and the Upper Amazon basin (Fig. 1a). A DEM of the study area was extracted from the global DEM GTOPO30 (U.S. Geological Survey, 1996). The DEM includes 230,880 points (the matrix 480 × 481); the grid spacing is 30" (Fig. 1a).

This area and GTOPO30 were selected because this DEM incorporates a high-frequency noise caused by interpolation errors and inaccurate merging of topographic charts having different accuracy. Spatial distribution of the noise in GTOPO30 is uneven and depends on the accuracy of cartographic sources. In particular, the potent noise is typical for forested regions of South America because reasonably detailed and accurate topographic data were unavailable for such areas. Thus, interpolation of sparse contours has been used to compile these portions of GTOPO30. Although DEM noise is no obstacle to producing realistic maps of elevation, it leads to derivation of noisy and unreadable maps of local morphometric variables (computation of the first and second partial derivatives of elevation dramatically increases the noise – Florinsky, 2002). The study area, consisting of two main zones – high mountains and forested foothills – which have different signal-to-noise ratio, is ideally suited to validate the method and algorithm proposed as a tool for DEM analytical treatment. We earlier used this DEM to evaluate two-dimensional singular spectrum analysis as a tool for filtering of digital terrain models (Golyandina et al., 2007).

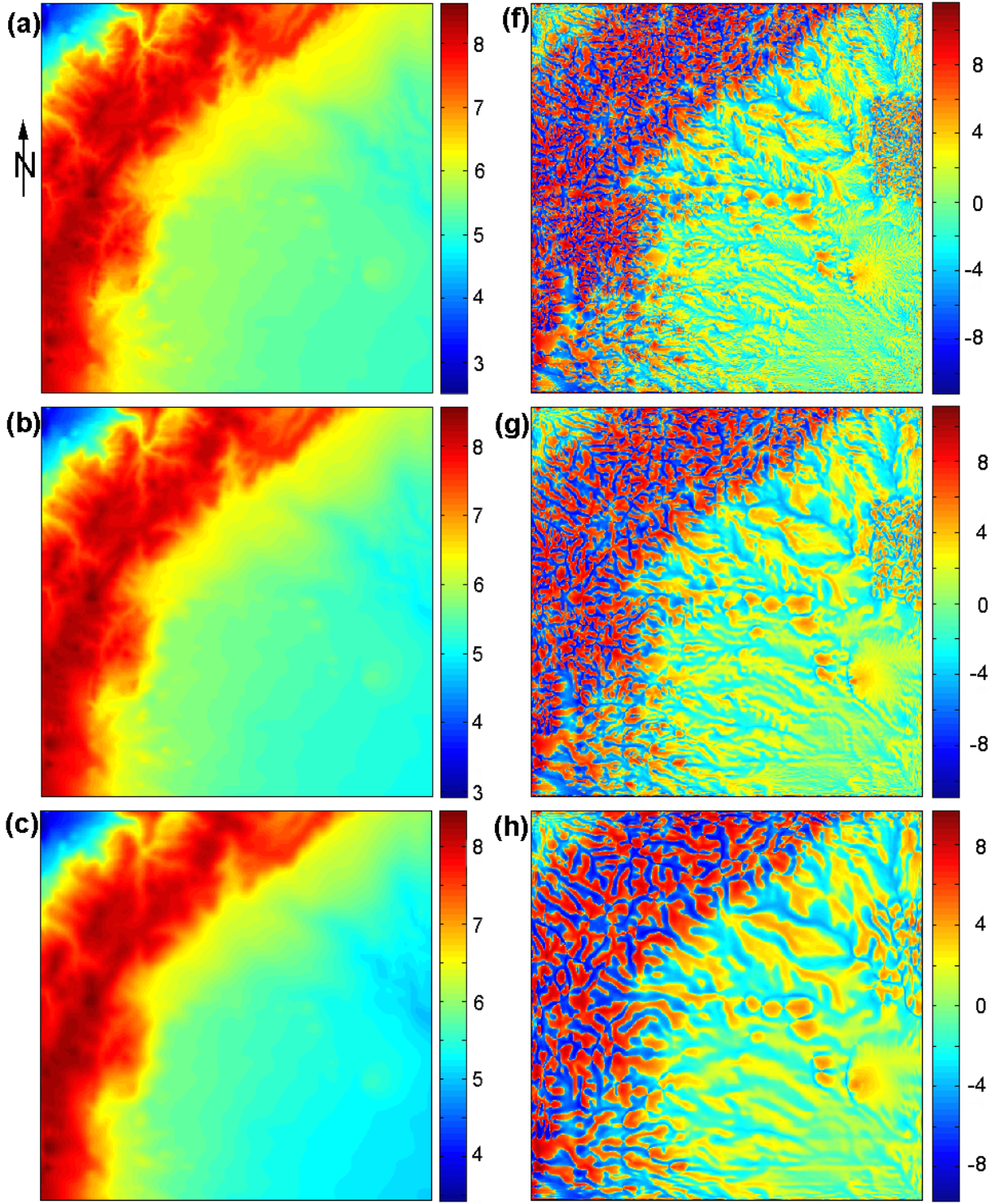

Fig. 1. The Northern Andes: elevations reconstructed with: (a) 480, (b) 240, (c) 120, (d) 60, and (e) 30 expansion coefficients; $k_h$ derived from reconstructed DEMs with: (f) 480, (g) 240, (h) 120, (i) 60, and (j) 30 expansion coefficients. Map legends are in logarithmic scale.

## 5. Data processing

We evaluated various numbers of expansion coefficients to reconstruct DEMs. Finally, the most expressive variants were selected to illustrate capabilities of the method and algorithm to generalize and denoise DEMs as well as to calculate partial derivatives for further computation of local morphometric attributes. In particular, DEMs were reconstructed with 480, 240, 120, 60, and 30 expansion coefficients (Fig. 1a, b, c, d, e). The first and second partial derivatives were analytically calculated from the reconstructed DEMs. Digital models of several local topographic attributes were then computed using the derivatives. To illustrate efficiency of the method and algorithm, we present $k_h$ maps (Fig. 1f, g, h, i, j).

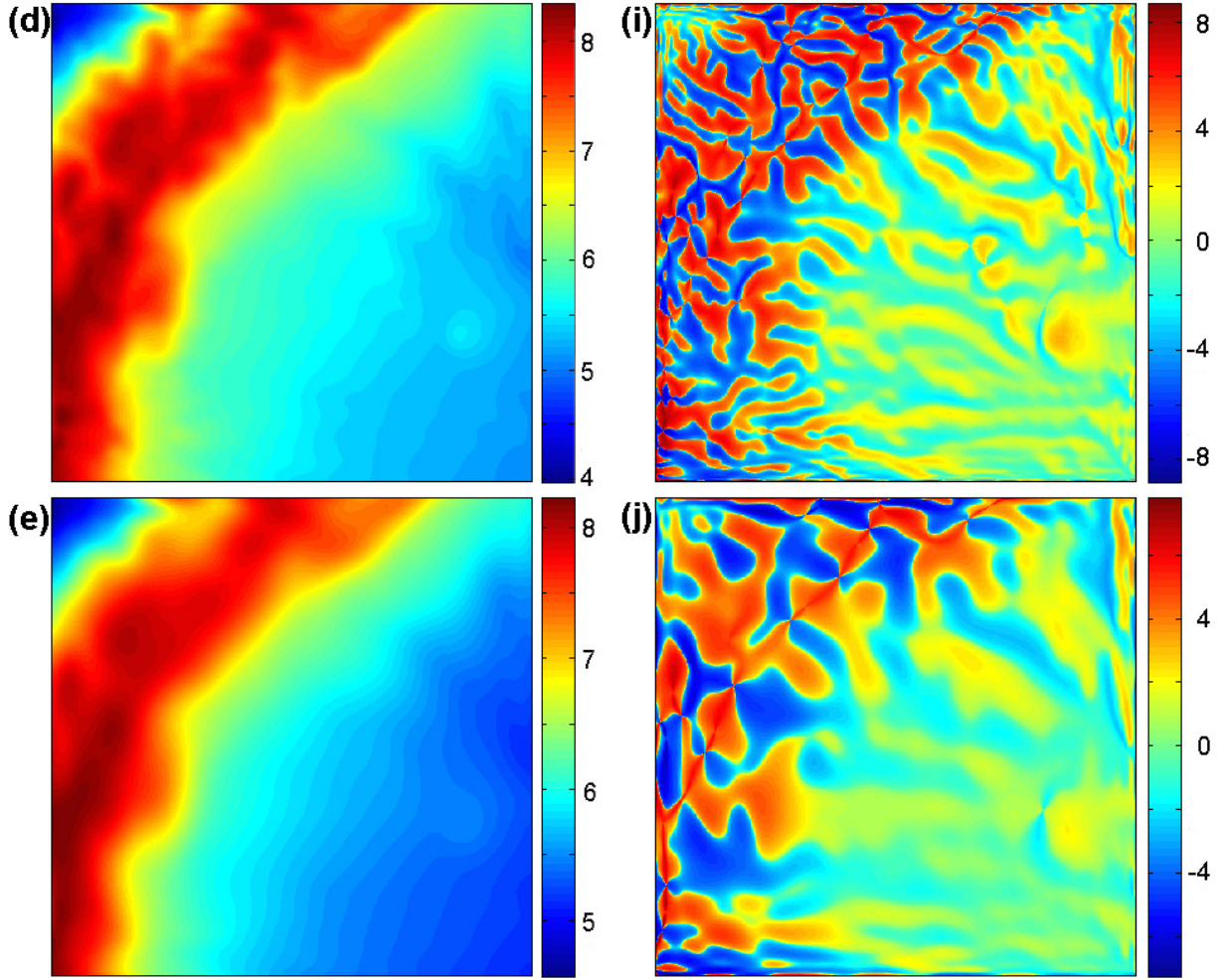

Fig. 1. (Cont.inued).

Wide dynamic ranges usually characterize topographic variables. For example, the range of elevations within the study is about 6080 m. To avoid loss of information on spatial distribution of values of morphometric attributes in mapping, it makes sense to apply a logarithmic transform using the following expression (Shary et al., 2002; Florinsky, 2012, p. 134):

$$\Theta' = \text{sign}(\Theta)\ln(1 + 10^n|\Theta|), \qquad (17)$$

where $\Theta$ is a value of a morphometric variable, $n = 0$ for elevation and nonlocal variables and $n = 2, \ldots, 18$ for local variables. For the DEM analytical treatment, selection of the $n$ value depends on the size of a study area (we used $n = 8$ for $k_h$ mapping). Such a form of logarithmic transformation considers that dynamic ranges of some topographic attributes include both positive and negative values. All DEMs reconstructed and digital models of $k_h$ derived were logarithmically transformed to prepare readable maps of elevation and $k_h$ (Fig. 1).

DTMs produced had a grid size of 30″. The plate carree projection was used for mapping. Data processing and mapping was done by software Matlab R2008b.

## 6. Results and discussion

A set of elevation maps reconstructed with 480, 240, 120, 60, and 30 expansion coefficients (Fig. 1a–e) demonstrates a process of DEM generalization, from its minimal level (Fig. 1a) to the maximal one (Fig. 1e). Similarly, a set of $k_h$ maps calculated from DEMs reconstructed with 480, 240, 120, 60, and 30 expansion coefficients (Fig. 1f–j) shows $k_h$ generalization, from its minimal level (Fig. 1f) to the maximal one (Fig. 1j).

It is important to note that a visual comparison of elevation maps reconstructed with 480, 240, 120 expansion coefficients (Fig. 1a, b, c) allows one to see nothing but marginal changes in image patterns. A cursory examination may lead to an underestimation of results of the DEM generalization. $k_h$ maps give better insight into the results. A comparison between $k_h$ maps derived from different DEMs (Fig. 1f–j) shows a pronounced effect of the map generalization. The less number of the expansion coefficients used to reconstruct a DEM, the more smooth and simplified image patterns obtained. One can see so-called flow structures formed by convergence and divergence areas (negative and positive $k_h$ values, correspondingly).

The DEMs reconstructed with 60 and 30 expansion coefficients (Fig. 1d, e) are marked by the highest level of generalization. These elevation maps represent a generalized morphostructure of the continental scale, the Andean Range with foothills.

The decrease of the number of the expansion coefficients in DEM reconstruction acts as a high frequency filtering. Indeed, manifestation of high frequency noise can be found on the $k_h$ map derived from the DEM reconstructed with 480 expansion coefficients (see the bottom right corner in Fig. 1f). This noise is typical for the Andean foothills covered by dense rain forests. However, there are no traces of this noise on $k_h$ maps derived from DEMs reconstructed with less number of expansion coefficients (Fig. 1g–j). One can also observe the removal of an artifact of other sort – a rectangular feature along the northeastern border of several $k_h$ maps (Fig. 1f–h). This is a trace of the inaccurate merging of adjacent topographic charts marked by different accuracy during the compilation of GTOPO30. All traces of the artifact disappear on $k_h$ maps calculated from DEMs reconstructed with 60 and 30 expansion coefficients (Fig. 1i–j).

It is well known that by orthogonal polynomial approximation produces boundary effects, which cannot be completely eliminated. In our case, they appear as linear artifacts at the boundaries of $k_h$ maps (Fig. 1f–h). A problem with map boundaries also arises using the finite difference algorithms: it is impossible to estimate partial derivatives for boundary columns and rows of a DEM, because they are estimated for the center point of the moving window (Evans, 1979, p. 29; Zevenbergen and Thorne, 1987; Shary, 1995; Florinsky, 1998, 2009).

Note that the above mentioned noise and artifacts are weakly expressed: they are not visible on the elevation maps (Fig. 1a–e) and become clearly visible only on $k_h$ maps (Fig. 1f–j) after the calculation of partial derivatives of elevation, which reinforce their expression.

Processing time to approximate a DEM and calculate analytically a morphometric variable is less than processing time to calculate a morphometric variable by a finite difference algorithm. So, the efficiency of the algorithm developed is higher than that of existing finite difference algorithms.

## 7. Conclusions

The test results demonstrated a good performance of the developed method and algorithm. They can be utilized as a universal tool for analytical treatment of DEMs including DEM global approximation, denoising, and generalization as well as derivation of local morphometric variables from DEMs using analytical calculation of partial derivatives. Further development of the method will include: (a) incorporation of other orthogonal polynomials (e.g., the Fourier and Legendre ones); (b) their comparative analysis in terms of efficiency and accuracy of DEM treatment; (c) evaluation of calculation accuracy using common statistical metrics (e.g., root mean square error of a function of measured values); and (d) comparison of the developed method and algorithm with traditional techniques applied in digital terrain modeling. We believe that it will be finally possible to avoid traditional methods of DEM interpolation and filtering as well as – what is more important – calculation of the morphometric characteristics using finite difference algorithms.

## Acknowledgements

The study was supported by RFBR grant 15-07-02484.